\documentclass[apjl]{emulateapj}
\usepackage{apjfonts}

\def\wig#1{\mathrel{\hbox{\hbox to 0pt{%
          \lower.5ex\hbox{$\sim$}\hss}\raise.4ex\hbox{$#1$}}}}

\shorttitle{Planetary Spectra: Evidence for Water Vapor?}
\shortauthors{Fortney \& Marley}

\newcommand{\hd}{HD~209458b}

\newcommand{\he}{HD~189733b}

\slugcomment{To be published in ApJ Letters}

\begin{document}

\title{Analysis of \emph{Spitzer} Spectra of Irradiated Planets: Evidence for Water Vapor?}

\author{Jonathan J. Fortney\altaffilmark{1}$^,$\altaffilmark{2} \& Mark S. Marley\altaffilmark{1}}

\altaffiltext{1}{Space Science \& Astrobiology Division, NASA Ames Research Center, Moffett Field, CA 94035; jfortney@arc.nasa.gov, mark.s.marley@nasa.gov}
\altaffiltext{2}{Spitzer Fellow, Carl Sagan Center, SETI Institute, 515 North Whisman Road, Mountain View, CA 94043}

\begin{abstract}
Published mid infrared spectra of transiting planets \hd~and \he, obtained during secondary eclipse by the InfraRed Spectrograph (IRS) aboard the \emph{Spitzer Space Telescope}, are predominantly featureless.  In particular these flux ratio spectra do not exhibit an expected feature arising from water vapor absorption short-ward of $10\,\rm \mu m$.  Here we suggest that, in the absence of flux variability, the spectral data for \he~are inconsistent with  $8\,\mu$m-photometry obtained with  \emph{Spitzer}'s InfraRed Array Camera (IRAC), perhaps an indication of problems with the challenging reduction of the IRS spectra.  The IRAC point, along with previously published secondary eclipse photometry for \he, are in good agreement with a one-dimensional model of \he~that clearly shows absorption due to water vapor in the emergent spectrum. We are not able to draw firm conclusions regarding the IRS data for \hd, but spectra predicted by 1D and 3D atmosphere models fit the data adequately, without adjustment of the water abundance or reliance on cloud opacity.  We argue that the generally good agreement between model spectra and IRS spectra of brown dwarfs with atmospheric temperatures similar to these highly irradiated planets lends confidence in the modeling procedure.

\end{abstract}


\clearpage

\section{Introduction}
Since the detection of the first extrasolar giant planet (EGP), 51 Peg b \citep{Mayor95}, considerable effort has gone into both observing and modeling properties of the atmospheres of the close-in ``hot Jupiter'' planets.  The \emph{Spitzer Space Telescope} now provides us a unique probe into the mid-infrared emission from these exotic atmospheres.

In this Letter we examine mid infrared spectra of transiting planets \hd~and \he, which were recently observed with \emph{Spitzer}'s InfraRed Spectrograph (IRS) by \citet{Richardson07} and \citet{Grillmair07}, respectively.  The spectra were obtained from $\sim$7.5 to 14\,$\mu$m around the time of secondary eclipse.  The most prominent conclusion of these works is that strong absorption arising from H$_2$O, expected short-ward of 10\,$\mu$m, was not seen for either planet.  Possible explanations put forward for the relatively featureless spectra included a lack of H$_2$O in these atmospheres, masking by a continuum of opaque clouds, or an isothermal temperature structure.

Recently \citet{Knutson07b} published an 8\,$\mu$m light curve for \he\ that includes an observation of the secondary eclipse.  These observations used \emph{Spitzer}'s InfraRed Array Camera (IRAC).  We suggest that this data point is inconsistent with the IRS spectrum, that water vapor is present and absorbs from 6.5-10\,$\mu$m, and that the published, low S/N spectrum of \he\ may not reflect the true planet spectrum.  We have computed model atmospheres and spectra of \he\ and \hd\ in an effort to shed additional light on these exciting observations.  We will also briefly discuss L- and T-type brown dwarfs, whose atmospheres span the same temperature range as hot Jupiters.  We suggest that the generally very good fits of our models to these high signal-to-noise IRAC and IRS datasets supports the veracity of the underlying physics and chemistry of our planet models.

\section{Spitzer Spectra of Planets}
Two teams employed the IRS instrument in ``Short-Low'' configuration to obtain low resolution spectra of parent stars HD~209458 and HD~189733 around the time of planetary secondary eclipse.  Each group compared the spectrum of the system when the planet's day side was still visible (before and after the secondary eclipse) to the stellar spectrum when the planet was behind the star.  The observed enhancement in flux seen when the planet was visible was attributed to the planet, thus providing a measurement of the planet-to-star flux ratio.  These exciting and difficult observations are at the leading edge of spectral characterization of exoplanets.

\begin{figure*}
\includegraphics[scale=0.82,angle=90]{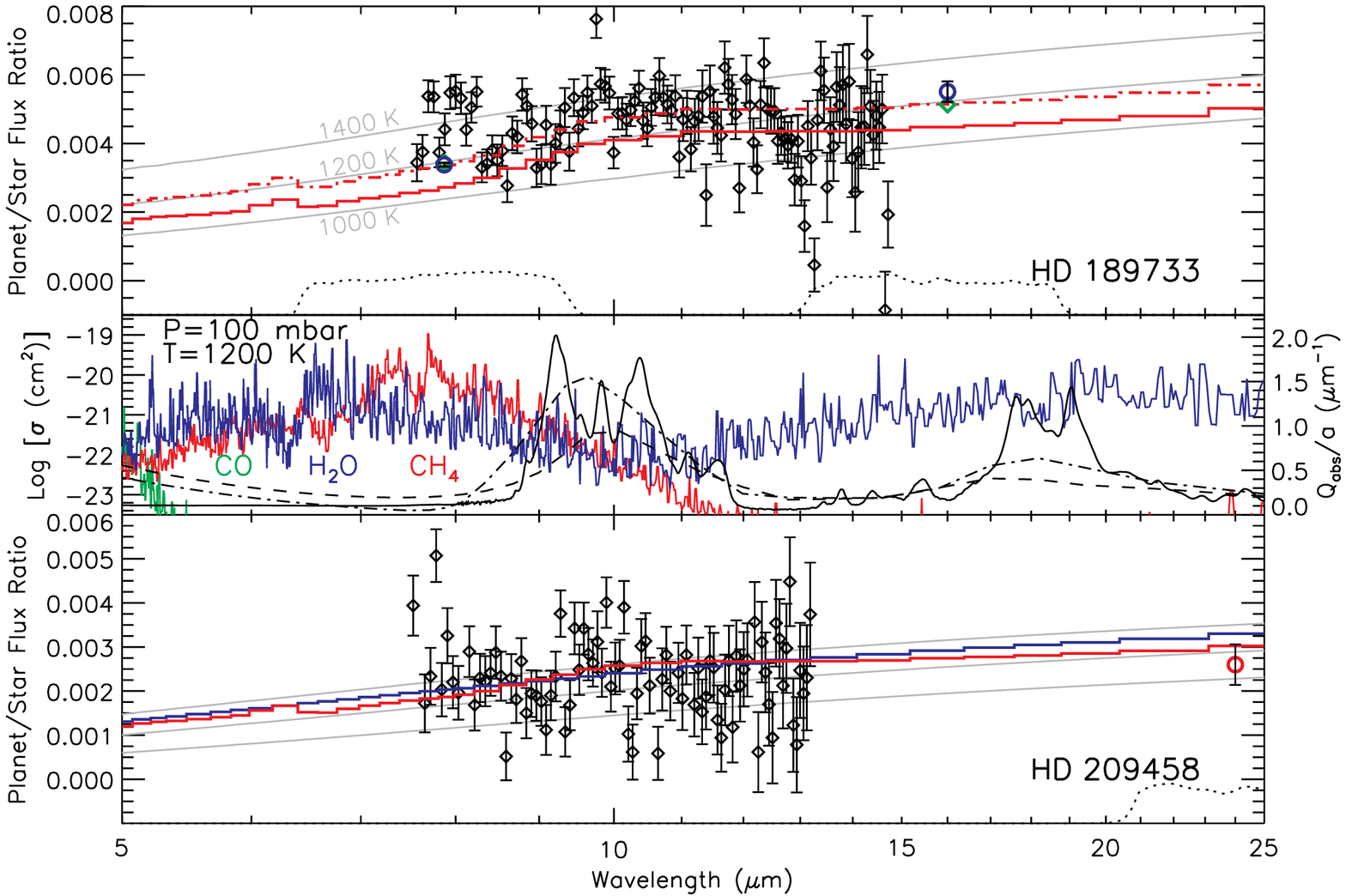}
\caption{(\emph{a}) Planet-to-star flux ratio for planet \he.  Black diamonds are the IRS secondary eclipse data from \citet{Grillmair07}.  The blue circles are secondary eclipse data from \citet{Knutson07b} at 8\,$\mu$m and \citet{Deming06} at 16\,$\mu$m.  All error bars are 1$\sigma$.  The dotted curves are transmission functions for these photometric bandpasses.  In gray are flux ratios for blackbody planetary emission at 1400, 1200, and 1000 K.  The red solid curve is a 1D uniform day/night model.  The red dash-dot curve is a model with a hotter day side.  Green diamonds are computed ratios for the dash-dot model across the 8 and 16\,$\mu$m bands.  (\emph{b}) Shown are absorption cross-sections (left axis) for CO (green), H$_2$O (blue), and CH$_4$ (red).  In black are the opacities (Q$_{abs}$/particle radius, right axis) of 1\,$\mu$m radius spheres of amorphous forsterite (dashed), amorphous enstatite (dash-dot), and crystalline enstatite (solid).  (\emph{c}) Planet-to-star flux ratio for planet \hd.  Black diamonds are the IRS secondary eclipse data from \citet{Richardson07}.  The red circle is the secondary eclipse observation from \citet{Deming05b} at 24\,$\mu$m.  The dotted curve is the blue edge of the transmission function for this bandpass. The gray curves are the planetary blackbodies from above.  The red curve is from the 1D uniform day/night model.  The blue curve is from the day side of the 3D model from \citet{Fortney06b}.
\label{3pan}}
\end{figure*}

\citet{Grillmair07} observed two secondary eclipse events of planet \he~and from these observations published a planet-to-star flux ratio from 7.57 to 14.71\,$\mu$m.  They found a flux ratio spectrum that is essentially flat across these wavelengths with a mean eclipse depth of $0.0049\pm 0.0002$.  The subset of their data with the highest signal-to-noise is shown in \mbox{Figure~\ref{3pan}}\emph{a}.  \citet{Richardson07} observed two secondary eclipses of planet \hd~and published a flux ratio spectrum from 7.5 to 13.2\,$\mu$m.  The derived average ratio is shown in \mbox{Figure~\ref{3pan}}\emph{c}.  These authors also find a spectrum that could be characterized as mostly flat.  However, they also detect increases in the flux ratio from 9.3-10\,$\mu$m, as well as a narrow emission feature at 7.7\,$\mu$m, which they attribute to emission from a stratosphere.

In addition to spectra, we plot \emph{Spitzer} secondary eclipse photometric data.  These points include the IRS blue peak-up at 16\,$\mu$m \citep{Deming06} and the IRAC 8\,$\mu$m point \citep{Knutson07b} for \he, and the Multiband Imaging Photometer for Spitzer (MIPS) point at 24\,$\mu$m for \hd~\citep{Deming05b}.  We note that in \mbox{Figure~\ref{3pan}}\emph{a} the IRAC 8\,$\mu$m point clearly plots below the IRS data, which we'll discuss in detail in \S2.2.

\subsection{Model Atmospheres and Spectra}
In order to further the study of these observations and their constraints on the character of each planet's atmosphere we have computed model atmospheres and generated low resolution spectra.  We employ a 1D model atmosphere code that has been used for a variety of planetary and substellar objects.  Recently it has been used for brown dwarfs \citep{Marley02,Saumon06} and EGPs \citep{Fortney05,Fortney06,Marley07b}.  It explicitly includes both incident radiation from the parent star (if present) and thermal radiation from the planet's atmosphere.  The radiative transfer solving scheme was developed by \citet{Toon89} and was first applied to irradiated giant planet atmospheres by \citet{Marley99}.  We use the elemental abundance data of \citet{Lodders03} and compute chemical equilibrium compositions following Lodders \& Fegley (2002, 2006).  Our opacity database is described in \citet{Freedman07}.  The stellar models are from \citet{Kurucz93} and the system parameters are from \citet{Winn07} for \he~and \citet{Knutson07a} for \hd.  The planet models presented here are cloud-free.

We plot model planet-to-star flux ratios in \mbox{Figure~\ref{3pan}}.  For \he~we show two one-dimensional models.  The solid red curve is for a day-side model that assumes all incident flux is efficiently redistributed over the entire planet, meaning that the incident flux is multiplied by a factor of 1/4 \citep{Appleby84}.  Following conventions adopted in solar system atmosphere modeling \citep[e.g.][]{Mckay89}, here and in our previous papers \citep{Fortney05,Fortney06} the incident flux is decreased by incorporating a multiplicative factor $f$ of 1/2 due to the day-night average, and another factor of 1/2 arises from the mean 60$^\circ$ stellar zenith angle ($\mu$, the  cosine of this angle, is 0.5) that is correct for global average insolation conditions \citep{Mckay89,MM99}.  With this prescription more stellar flux is absorbed at low pressures, leading to a warmer upper atmosphere and shallower temperature gradient, compared to a model where incident flux is directly multiplied by 1/4 with $\mu$=1.0.  The greater absorption of incident flux at lower pressure is due to the twice longer path length for incident photons to reach a given pressure.  In reality, these differences in 1D models could be swamped by a more complex day-side temperature structure for these planets.  Here, for models with a warmer day side, the incident flux is multiplied by a factor $f$ between 1/2 and 1, while $\mu$ remains 0.5.  The dash-dot curve in \mbox{Figure~\ref{3pan}}\emph{c} is for a warmer model, where $f=4/5$, meaning that only 20\% of the absorbed incident flux is assumed to be advected to the night side.  Both models predict an absorption feature due to H$_2$O short-ward of 10\,$\mu$m. \mbox{Figure~\ref{3pan}}\emph{b} shows absorption cross sections for H$_2$O, CO, and CH$_4$.  At these temperatures ($\sim$1200 K), CO is strongly favored at the expense of CH$_4$.  The dashed model in \mbox{Figure~\ref{3pan}}\emph{a} is a good ($\sim$1$\sigma$) fit to the band-averaged 8\,$\mu$m \citep{Knutson07b} and 16\,$\mu$m \citep{Deming06} secondary eclipse observations.  We note that although \citet{Grillmair07} claim their flat ratio spectrum is consistent with a blackbody, \mbox{Figure~\ref{3pan}}\emph{a} shows that it is actually bluer than a blackbody.

\mbox{Figure~\ref{3pan}}\emph{c} shows the mean \citet{Richardson07} data set.  In solid red we plot the $f=1/2$ 1D model spectrum from \citet{Fortney05}.  Note that the predicted ``downturn'' in the ratio spectrum short-ward of 10\,$\mu$m is actually quite small in this 1D model, less than that predicted by other modelers \citep[e.g.,][]{Seager05,Burrows06}, for reasons that are not yet clear, but could be related to the angle of incident radiation.  Given the S/N of this dataset and the relatively small predicted depth of the model H$_2$O absorption feature, it is not clear that absorption due to H$_2$O can be dismissed.

Dynamical processes can act to produce isothermal profiles in the upper atmosphere which would also suppress the depths of absorption features.  In solid blue we plot the spectrum of the day side of the \citet{CS06} 3D dynamical model of \hd, as computed in \citet{Fortney06b}.  The isothermal atmosphere is due to a super-rotating west-to-east equatorial jet; as cool night side gas is blown around the planet's limb this gas is more readily warmed at lower pressures where the radiative time constant is shorter \citep{Iro05}, leading to day side \emph{P-T} profiles that are more isothermal than one would predict for a radiative equilibrium model \citep{Showman06}.   

Neither the 1D nor 3D model has the increase in flux ratio seen from $\sim9.3-10$\,$\mu$m, as neither model has the strong temperature inversion that would be necessary to create emission features.  In \mbox{Figure~\ref{3pan}}\emph{b} we plot the Mie absorption efficiency of 1\,$\mu$m silicate particles of forsterite (Mg$_2$SiO$_4$) and enstatite (MgSiO$_3$).  Larger particles have a grayer spectrum and do not exhibit the canonical silicate feature \citep{Hanner94}.  The location of this possible emission feature in the \hd~spectrum is suggestive, but it will take higher signal-to-noise observations to determine if silicates have indeed been detected.  \hd~is a better candidate than \he~for having silicate clouds in its visible atmosphere, due to the higher stellar insolation and hotter atmosphere.

\begin{figure}
\epsscale{1.2}
\plotone{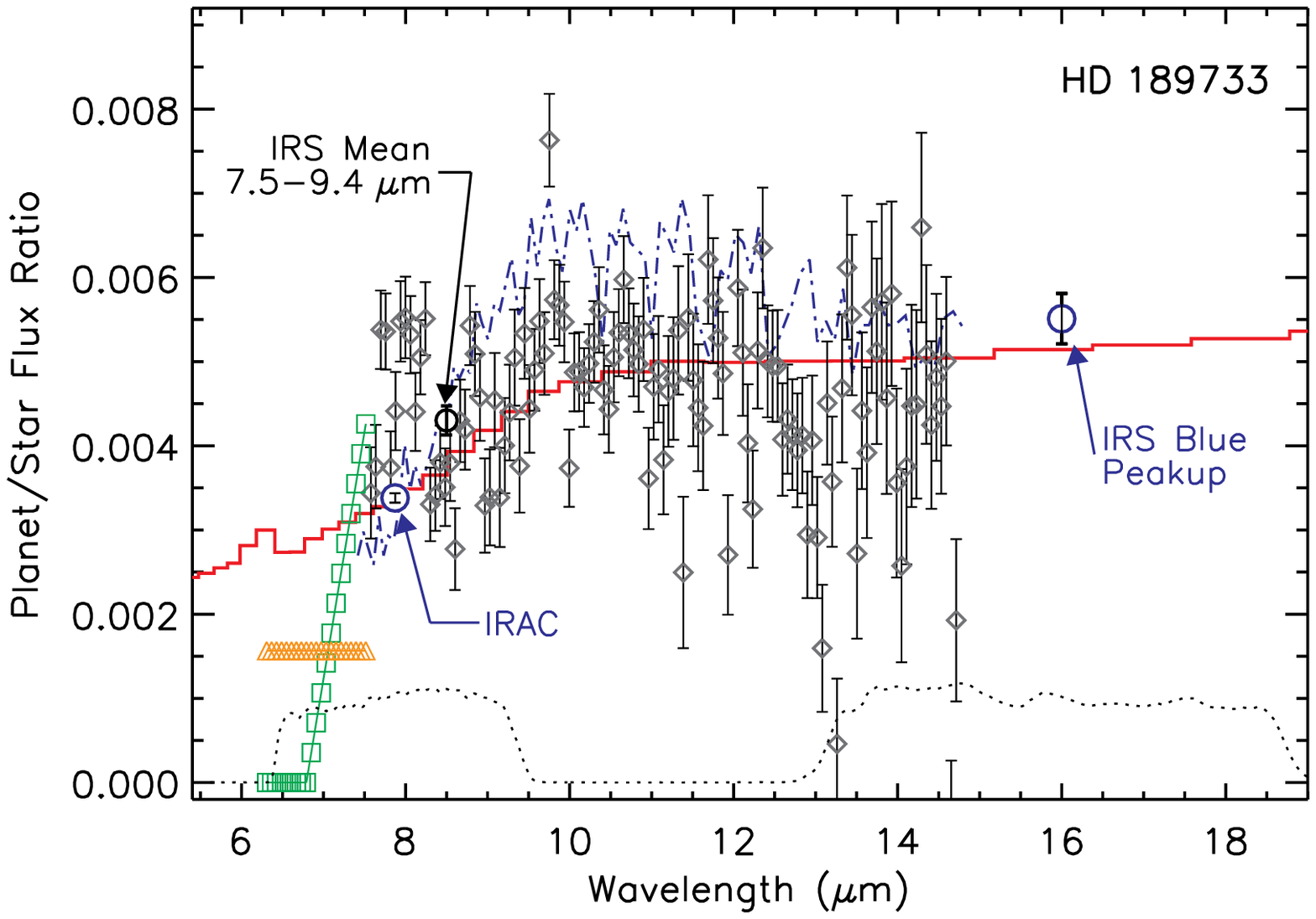}
\caption{Planet-to-star flux ratio for planet \he.  In diamonds are the \citet{Grillmair07} IRS data.  The blue circles are from secondary eclipse photometry.  The scaled band transmission functions are shown as dotted lines.  The black circle is the eclipse depth from 7.57 to 9.39\,$\mu$m from the IRS data.  All error bars are 1$\sigma$.  The triangles and squares show flux ratios short-ward of the IRS data that are small enough to bring the flux ratio across the entire IRAC band (6.3 to 9.4\,$\mu$m) into precise agreement with the IRAC observation.  (See text.)  In solid red is our $f=4/5$ model and dash-dot blue is the \citet{Burrows06} model as shown in \citet{Grillmair07}.
\label{189}}
\end{figure}

\subsection{A Closer Look at \he}
As was noted above and shown in \mbox{Figure~\ref{3pan}}\emph{a}, the IRAC observation at 8\,$\mu$m by \citet{Knutson07b} indicates a downturn that is in fact consistent with H$_2$O vapor absorption.  We now take a closer look in \mbox{Figure~\ref{189}}.  Our model (red) is a good fit to the data, while in blue we show a \citet{Burrows06} model, as taken from \citet{Grillmair07}, that appears to also be a good fit to the photometry, albeit with a much deeper water feature.  Binning the 31 IRS data points from 7.57 to 9.39\,$\mu$m (which overlaps the IRAC band) into a single point, we find an eclipse depth of $0.00430\pm0.00017$, $\sim$5.3$\sigma$ removed from the Knutson et al.~IRAC point.  If variation in planetary or stellar flux were to account for this difference \citep[e.g.][]{Menou03,Rauscher07}, the variability would have to be $\sim$25\% near 8\,$\mu$m.  If there is no variability and both the IRS spectrum and IRAC data point are accurate as published, we can examine how small the flux ratio must be short-ward of the bluest IRS point at 7.57\,$\mu$m.  We have created synthetic data with the same spectral resolution of the IRS data set.  In triangles we show a constant flux ratio of 0.00156 between 6.3\,$\mu$m (the blue end of the IRAC filter) and 7.57\,$\mu$m.  In boxes we show a synthetic ratio that decreases linearly from the 0.00430 binned value.  The ratio must fall to \emph{zero} by 6.79\,$\mu$m to give a band-average eclipse depth that agrees precisely with the IRAC data point.

In \mbox{Figure~\ref{3pan}}\emph{b} we showed that the only potential absorbers predicted from equilibrium chemistry from 6 to 10\,$\mu$m are H$_2$O and CH$_4$, neither of which possess a sharp increase in opacity short-ward of 7.7\,$\mu$m.  With the assumption that the IRAC secondary eclipse depth is accurate, and given the severe absorption feature that would be needed just short of 7.57\,$\mu$m to fit this point given the IRS spectrum, we conclude that the published IRS spectrum short-ward of 10\,$\mu$m does not reflect the true spectrum at these wavelengths, where the IRAC data point suggests absorption by water vapor.

One possibility toward reconciling the data sets would be to shift the entire IRS spectrum downward.  Indeed, for \hd, \citet{Richardson07} were unable to accurately determine the absolute depth of the secondary eclipse with the IRS, and therefore tied their spectrum to an unpublished IRAC 8\,$\mu$m observation of the secondary eclipse.  Given this issue, the absolute depths of the \citet{Grillmair07} data may also require a shift.  However, the downward shift that would be required ($\sim$25\%) would lead to much poorer agreement between this data set and the \citet{Deming06} point at 16\,$\mu$m.  In addition, although both teams corrected for slit losses, this was only done in a gray manner.  Since the point spread function is larger at longer wavelengths, it is possible that slit losses would be more pronounced at the red end of the spectrum.  Greater flux at longer wavelengths, together with a downward scaling, may be able to bring the observed spectrum for \he~into agreement with the photometric points.  Future work on all of these data sets may be beneficial, as, for example, the relative photometry for these measurements is more than an order of magnitude more precise than IRAC's initial specifications \citep{Fazio04}.

\section{Lessons from Brown Dwarfs}
The \emph{P-T} conditions encountered by the atmospheres of hot Jupiters is coincident with that found in the atmospheres of L- and T-type brown dwarfs.  Although the actual \emph{P-T} profiles of irradiated EGPs are of course different, the underlying atmospheric physics and chemistry are similar and indeed the actual computational models are in some cases virtually identical, such that a comparison of brown dwarf models to data can illuminate theoretical areas that are well understood.

Comparisons of our group's models to near- and mid-infrared spectra of L- and T-dwarfs, including superb observations with the IRS, have generally found very good to excellent matches with data \citep{Golim04,Knapp04,Cushing06,Saumon06,Mainzer07,Leggett07}.  Specifically our ``cloudy'' L-dwarf and ``clear'' T dwarf models \citep[][M.~S. Marley et al. in prep.]{Marley02} fit the depths of the water bands lying between Y, J, H, and K bands as well as the $6.5\,\rm \mu m$ water band \citep{Cushing06, Saumon06, Mainzer07, Cushing07} expected in the IRS spectra of hot Jupiters.  There are, however, departures from the data in regions where molecular opacities are poorly known as well as some systematic spectral shape differences in K and N bands.  

Since the Marley et al.~models have been demonstrated to fit the near and mid infrared data of the cloud-free T dwarfs we gain confidence in our radiative transfer \citep{Toon89}, opacity database \citep{Freedman07}, and chemistry calculations \citep{Lodders02, Lodders06}.  This experience gives us confidence these areas are becoming well understood.  Of course, since isolated brown dwarfs lack incident flux that drive atmospheric dynamics and photochemistry, these additional areas may remain challenging topics of hot Jupiter characterization for some time.  It is in deviations from 1D or simple day/night models that these aspects will become apparent. 

\section{Discussion \& Conclusions}
The common conclusion of the \citet{Grillmair07} and \citet{Richardson07} secondary eclipse observations was the apparent lack of absorption due to water vapor.  However, for \he~we have shown that photometry indicates a planet-to-star flux ratio that is not as featureless as indicated by the published IRS data.  If variability in planetary flux is to make up for this discrepancy, it must reach $\sim$25\%.  The apparent downturn in the flux ratio from 16 to 8\,$\mu$m implied by photometry is consisent with a model that predicts absorption due to water shortward of 10\,$\mu$m.  As was shown in \mbox{Figure~\ref{189}}, there is considerable diversity in feature depths among models, perhaps due to differences in chemistry, opacities, or treatment of incoming stellar flux.

For \he, here we have shown that our model is a good match to 8 and 16\,$\mu$m-band photometry.  In addition, both our 1D and 3D models fit the \hd~observations reasonably well.  Our conclusion that water vapor absorption is present in these atmospheres is supported by the recent analysis of the \citet{Knutson07a} transit data set by \citet{Barman07}, who concluded that water absorption in present in the atmosphere of \hd~in a band at $\sim950$ nm.  Of course the characterization of the atmospheres of these planets is just beginning.  Both planets have been observed in all 4 IRAC bands, but this data has not yet been published.   Additional IRS spectroscopic observations for \hd~(Deming and collaborators) and \he~(Grillmair and collaborators) will be obtained this year.  Importantly, these observations will go down to 5.5\,$\mu$m, fully overlapping the IRAC 8\,$\mu$m band, which will allow for more accurate normalization of the obtained spectra. 
\\
\\
We thank J.~Richardson, M.~Cushing, D.~Deming, C.~Grillmair, D.~Charbonneau, and the referee for useful discussions.  J.~J.~F.~acknowledges the support of a Spitzer Fellowship from NASA and M.~S.~M.~from the NASA Origins and Planetary Atmospheres Programs.


\begin{thebibliography}{40}
\expandafter\ifx\csname natexlab\endcsname\relax\def\natexlab#1{#1}\fi

\bibitem[{{Appleby} \& {Hogan}(1984)}]{Appleby84}
{Appleby}, J.~F. \& {Hogan}, J.~S. 1984, Icarus, 59, 336

\bibitem[{{Barman}(2007)}]{Barman07}
{Barman}, T. 2007, \apjl, 661, L191

\bibitem[{{Burrows} {et~al.}(2006){Burrows}, {Sudarsky}, \&
  {Hubeny}}]{Burrows06}
{Burrows}, A., {Sudarsky}, D., \& {Hubeny}, I. 2006, \apj, 650, 1140

\bibitem[{{Cooper} \& {Showman}(2006)}]{CS06}
{Cooper}, C.~S. \& {Showman}, A.~P. 2006, \apj, 649, 1048

\bibitem[{{Cushing} {et~al.}(2006){Cushing}, {Roellig}, {Marley}, {Saumon},
  {Leggett}, {Kirkpatrick}, {Wilson}, {Sloan}, {Mainzer}, {Van Cleve}, \&
  {Houck}}]{Cushing06}
{Cushing}, M.~C., et al. 2006, \apj, 648, 614

\bibitem[{{Cushing} {et~al.}(2007){Cushing}, {Roellig}, {Marley}, {Saumon},
  {Leggett}, {Van Cleve}, \&
  {Houck}}]{Cushing07}
{Cushing}, M.~C., et al. 2007, \apj, submitted

\bibitem[{{Deming} {et~al.}(2006){Deming}, {Harrington}, {Seager}, \&
  {Richardson}}]{Deming06}
{Deming}, D., {Harrington}, J., {Seager}, S., \& {Richardson}, L.~J. 2006,
  \apj, 644, 560

\bibitem[{{Deming} {et~al.}(2005){Deming}, {Seager}, {Richardson}, \&
  {Harrington}}]{Deming05b}
{Deming}, D., {Seager}, S., {Richardson}, L.~J., \& {Harrington}, J. 2005,
  Nature, 434, 740

\bibitem[{{Fazio} {et~al.}(2004){Deming}, {Seager}, {Richardson}, \&
  {Harrington}}]{Fazio04}
{Fazio}, G.~G., et al.  2004, ApJS, 154, 10
  
\bibitem[{{Fortney} {et~al.}(2006{\natexlab{a}}){Fortney}, {Cooper}, {Showman},
  {Marley}, \& {Freedman}}]{Fortney06b}
{Fortney}, J.~J., {Cooper}, C.~S., {Showman}, A.~P., {Marley}, M.~S., \&
  {Freedman}, R.~S. 2006{\natexlab{a}}, \apj, 652, 746

\bibitem[{{Fortney} {et~al.}(2005){Fortney}, {Marley}, {Lodders}, {Saumon}, \&
  {Freedman}}]{Fortney05}
{Fortney}, J.~J., {Marley}, M.~S., {Lodders}, K., {Saumon}, D., \& {Freedman},
  R. 2005, \apjl, 627, L69

\bibitem[{{Fortney} {et~al.}(2006{\natexlab{b}}){Fortney}, {Saumon}, {Marley},
  {Lodders}, \& {Freedman}}]{Fortney06}
{Fortney}, J.~J., {Saumon}, D., {Marley}, M.~S., {Lodders}, K., \& {Freedman},
  R.~S. 2006{\natexlab{b}}, \apj, 642, 495

\bibitem[{{Freedman} {et~al.}(2007){Freedman}, {Marley}, \&
  {Lodders}}]{Freedman07}
{Freedman}, R.~S., {Marley}, M.~S., \& {Lodders}, K. 2007, Submitted to ApJS

\bibitem[{{Golimowski} {et~al.}(2004){Golimowski}, {Leggett}, {Marley}, {Fan},
  {Geballe}, {Knapp}, {Vrba}, {Henden}, {Luginbuhl}, {Guetter}, {Munn},
  {Canzian}, {Zheng}, {Tsvetanov}, {Chiu}, {Glazebrook}, {Hoversten},
  {Schneider}, \& {Brinkmann}}]{Golim04}
{Golimowski}, D.~A., et al. 2004, \aj, 127, 3516

\bibitem[{{Grillmair} {et~al.}(2007){Grillmair}, {Charbonneau}, {Burrows},
  {Armus}, {Stauffer}, {Meadows}, {Van Cleve}, \& {Levine}}]{Grillmair07}
{Grillmair}, C.~J., {Charbonneau}, D., {Burrows}, A., {Armus}, L., {Stauffer},
  J., {Meadows}, V., {Van Cleve}, J., \& {Levine}, D. 2007, \apjl, 658, L115

\bibitem[{{Hanner} {et~al.}(1994){Hanner}, {Hackwell}, {Russell}, \&
  {Lynch}}]{Hanner94}
{Hanner}, M.~S., {Hackwell}, J.~A., {Russell}, R.~W., \& {Lynch}, D.~K. 1994,
  Icarus, 112, 490

\bibitem[{{Iro} {et~al.}(2005){Iro}, {Bezard}, \& {Guillot}}]{Iro05}
{Iro}, N., {Bezard}, B., \& {Guillot}, T. 2005, \aap, 436, 719

\bibitem[{{Knapp} {et~al.}(2004){Knapp}, {Leggett}, {Fan}, {Marley}, {Geballe},
  {Golimowski}, {Finkbeiner}, {Gunn}, {Hennawi}, {Ivezi{\'c}}, {Lupton},
  {Schlegel}, {Strauss}, {Tsvetanov}, {Chiu}, {Hoversten}, {Glazebrook},
  {Zheng}, {Hendrickson}, {Williams}, {Uomoto}, {Vrba}, {Henden}, {Luginbuhl},
  {Guetter}, {Munn}, {Canzian}, {Schneider}, \& {Brinkmann}}]{Knapp04}
{Knapp}, G.~R., et al. 2004, \aj, 127, 3553

\bibitem[{{Knutson} {et~al.}(2007{\natexlab{a}}){Knutson}, {Charbonneau},
  {Allen}, {Fortney}, {Agol}, {Cowan}, {Showman}, {Cooper}, \&
  {Megeath}}]{Knutson07b}
{Knutson}, H.~A., et al. 2007{\natexlab{a}}, \nat, 447, 183

\bibitem[{{Knutson} {et~al.}(2007{\natexlab{b}}){Knutson}, {Charbonneau},
  {Noyes}, {Brown}, \& {Gilliland}}]{Knutson07a}
{Knutson}, H.~A., {Charbonneau}, D., {Noyes}, R.~W., {Brown}, T.~M., \&
  {Gilliland}, R.~L. 2007{\natexlab{b}}, \apj, 655, 564

\bibitem[{{Kurucz}(1993)}]{Kurucz93}
{Kurucz}, R. 1993, CD-ROM 13, ATLAS9 Stellar Atmosphere Programs and 2 km/s
  Grid (Cambridge:SAO)

\bibitem[{{Leggett} {et~al.}(2007){Leggett}, {Saumon}, {Marley}, {Geballe},
  {Golimowski}, {Stephens}, \& {Fan}}]{Leggett07}
{Leggett}, S.~K., {Saumon}, D., {Marley}, M.~S., {Geballe}, T.~R.,
  {Golimowski}, D.~A., {Stephens}, D., \& {Fan}, X. 2007, \apj, 655, 1079

\bibitem[{{Lodders}(2003)}]{Lodders03}
{Lodders}, K. 2003, \apj, 591, 1220

\bibitem[{{Lodders} \& {Fegley}(2002)}]{Lodders02}
{Lodders}, K. \& {Fegley}, B. 2002, Icarus, 155, 393

\bibitem[{{Lodders} \& {Fegley}(2006)}]{Lodders06}
---. 2006, {Astrophysics Update 2} (Springer Praxis Books, Berlin: Springer,
  2006)

\bibitem[{{Mainzer} {et~al.}(2007){Mainzer}, {Roellig}, {Saumon}, {Marley},
  {Cushing}, {Sloan}, {Kirkpatrick}, {Leggett}, \& {Wilson}}]{Mainzer07}
{Mainzer}, A.~K., et al. 2007, \apj, 662, 1245

\bibitem[{{Marley} {et~al.}(2007){Marley}, {Fortney}, {Seager}, \&
  {Barman}}]{Marley07b}
{Marley}, M.~S., {Fortney}, J., {Seager}, S., \& {Barman}, T. 2007, in
  Protostars and Planets V, ed. B.~{Reipurth}, D.~{Jewitt}, \& K.~{Keil},
  733--747

\bibitem[{{Marley} {et~al.}(1999){Marley}, {Gelino}, {Stephens}, {Lunine}, \&
  {Freedman}}]{Marley99}
{Marley}, M.~S., {Gelino}, C., {Stephens}, D., {Lunine}, J.~I., \& {Freedman},
  R. 1999, \apj, 513, 879

\bibitem[{{Marley} \& {McKay}(1999)}]{MM99}
{Marley}, M.~S. \& {McKay}, C.~P. 1999, Icarus, 138, 268

\bibitem[{{Marley} {et~al.}(2002){Marley}, {Seager}, {Saumon}, {Lodders},
  {Ackerman}, {Freedman}, \& {Fan}}]{Marley02}
{Marley}, M.~S., {Seager}, S., {Saumon}, D., {Lodders}, K., {Ackerman}, A.~S.,
  {Freedman}, R.~S., \& {Fan}, X. 2002, \apj, 568, 335

\bibitem[{{Mayor} \& {Queloz}(1995)}]{Mayor95}
{Mayor}, M. \& {Queloz}, D. 1995, \nat, 378, 355+

\bibitem[{{McKay} {et~al.}(1989){McKay}, {Pollack}, \& {Courtin}}]{Mckay89}
{McKay}, C.~P., {Pollack}, J.~B., \& {Courtin}, R. 1989, Icarus, 80, 23

\bibitem[{{Menou} {et~al.}(2003){Menou}, {Cho}, {Seager}, \&
  {Hansen}}]{Menou03}
{Menou}, K., {Cho}, J.~Y.-K., {Seager}, S., \& {Hansen}, B.~M.~S. 2003, \apjl,
  587, L113

\bibitem[{{Rauscher} {et~al.}(2007){Rauscher}, {Menou}, {Cho}, {Seager}, \&
  {Hansen}}]{Rauscher07}
{Rauscher}, E., {Menou}, K., {Cho}, J.~Y.-K., {Seager}, S., \& {Hansen},
  B.~M.~S. 2007, \apjl, 662, L115

\bibitem[{{Richardson} {et~al.}(2007){Richardson}, {Deming}, {Horning},
  {Seager}, \& {Harrington}}]{Richardson07}
{Richardson}, L.~J., {Deming}, D., {Horning}, K., {Seager}, S., \&
  {Harrington}, J. 2007, \nat, 445, 892

\bibitem[{{Saumon} {et~al.}(2006){Saumon}, {Marley}, {Cushing}, {Leggett},
  {Roellig}, {Lodders}, \& {Freedman}}]{Saumon06}
{Saumon}, D., {Marley}, M.~S., {Cushing}, M.~C., {Leggett}, S.~K., {Roellig},
  T.~L., {Lodders}, K., \& {Freedman}, R.~S. 2006, \apj, 647, 552

\bibitem[{{Seager} {et~al.}(2005){Seager}, {Richardson}, {Hansen}, {Menou},
  {Cho}, \& {Deming}}]{Seager05}
{Seager}, S., {Richardson}, L.~J., {Hansen}, B.~M.~S., {Menou}, K., {Cho},
  J.~Y.-K., \& {Deming}, D. 2005, \apj, 632, 1122

\bibitem[{{Showman} \& {Cooper}(2006)}]{Showman06}
{Showman}, A.~P. \& {Cooper}, C.~S. 2006, in Tenth Anniversary of 51 Peg-b, ed. L.~{Arnold},
  F.~{Bouchy}, \& C.~{Moutou}, 242--250

\bibitem[{{Toon} {et~al.}(1989){Toon}, {McKay}, {Ackerman}, \&
  {Santhanam}}]{Toon89}
{Toon}, O.~B., {McKay}, C.~P., {Ackerman}, T.~P., \& {Santhanam}, K. 1989,
  Journal of Geophysical Research, 94, 16287

\bibitem[{{Winn} {et~al.}(2007){Winn}, {Holman}, {Henry}, {Roussanova}, {Enya},
  {Yoshii}, {Shporer}, {Mazeh}, {Johnson}, {Narita}, \& {Suto}}]{Winn07}
{Winn}, J.~N., et al. 2007, \aj, 133, 1828

\end{thebibliography}

\end{document}